\documentclass[10pt, twocolumn]{article}

\usepackage{amsmath,amssymb}
\usepackage{abstract}
\usepackage{graphicx}
\usepackage{caption}
\usepackage{color}
\usepackage{soul}
\usepackage{dcolumn}
\usepackage{bm}
\usepackage[numbers,comma,sort&compress]{natbib}
\usepackage[normalem]{ulem}
\usepackage{xurl}
\usepackage[colorlinks=true,linkcolor=blue!50!black,citecolor=blue!50!black,urlcolor=blue!50!black]{hyperref}

\definecolor{background-color}{gray}{0.98}

\usepackage{todonotes}

\usepackage[utf8]{inputenc}
\usepackage[T1]{fontenc}

\newcommand{\beq}{\begin{equation}}
\newcommand{\eeq}{\end{equation}}
\newcommand{\bqa}{\begin{eqnarray}}
\newcommand{\eqa}{\end{eqnarray}}
\newcommand{\nn}{\nonumber}

\newcommand{\smallfrac}[2]{\mbox{$\frac{#1}{#2}$}}

\newcommand{\half}{\smallfrac{1}{2}}

\newcommand{\tp}{^{\top}}

\newcommand{\cell}{\operatorname{Cell}}

\definecolor{maroon}{rgb}{0.7,0,0}
\definecolor{ngreen}{rgb}{0.2,0.6,0.2}
\definecolor{golden}{rgb}{0.8,0.6,0.1}

\newcommand{\blk}{\color{black}}

\title{Eigenstates in the Many Interacting Worlds approach: Ground states in 1D and 2D and excited states in 1D\\(long
version)}
\author{Hannes Herrmann\thanks{Mathematical Institute, LMU Munich, Germany},
    Michael J. W. Hall\thanks{Centre for Quantum Dynamics, Griffith University,
    Brisbane, QLD 4111, Australia}, Howard M. Wiseman\thanks{Centre for
    Quantum Dynamics, Griffith University, Brisbane, QLD 4111, Australia},
    Dirk - Andr\'e Deckert\thanks{Mathematical Institute, LMU Munich,
Germany}}

\makeatletter
\renewcommand\@biblabel[1]{#1.}
\makeatother

\begin{document}

\twocolumn[
\maketitle
\begin{@twocolumnfalse}
    \begin{abstract}
        Recently the Many-Interacting-Worlds (MIW) approach to a quantum theory
        without wave functions was proposed.  This \blk approach leads quite
        naturally to numerical integrators of the Schr\"{o}dinger equation.  It
        has been suggested that such integrators may feature advantages over
        fixed-grid methods for higher numbers of degrees of freedom.  However,
        as yet, \blk little is known about concrete MIW models for more than
        one spatial dimension and/or more than one particle. In this work we
        develop the MIW approach further to treat arbitrary degrees of freedom,
        and provide a systematic study of a corresponding numerical
        implementation for computing one-particle ground and excited states in
        one  dimension, \blk and ground states in two spatial dimensions.  With
        this step towards the treatment of higher degrees of freedom we hope to
        stimulate their further study.  
    \end{abstract}
    \vskip1cm
\end{@twocolumnfalse}
]

\normalsize
\saythanks


\section{Introduction}

The quantum dynamics of an $N$-particle system in $d$ spatial dimensions is ruled by the Schrödinger equation. The
latter defines the evolution of a field $\Psi$ on the configuration space $\mathbb R^{Nd}$. A common method to represent
such an object numerically is to sample it on a fixed grid. The grid dimensions scale as $G^{Nd}$,   with   $G$ being
the number of grid points along one degree of freedom. This exponential scaling behavior makes integrators based on
fixed grids, even on state-of-the-art supercomputers, unfeasible -- e.g., settings with $N=3$ particles in $d=3$
dimensions and a low number of $G=10^3$ grid points already require a memory capacity of the order of gigabytes.

Exceptional cases aside, the main \blk two successful general approaches to deal with this quantum complexity problem
are as follows. \blk 1) For special initial values and in certain regimes (e.g., product states and high gas densities)
one can find approximate solutions by solving non-linear one-particle equations such as the Hartree, Hatree-Fock and
Gross-Pitajevski equations; see, e.g., Ref.~\cite{lieb} \blk  for   an overview. 2) Instead of using a fixed equidistant
grid one may employ a comoving grid that samples the wave functions with high resolution only where it has physically
interesting features, while other regions are only covered with very few grid points; see Refs.~\cite{wyatt,chatteraj}
\blk for an overview. With a well-adapted grid it should in principle be possible to maintain the same accuracy with a
significantly lower number of grid-points, especially \blk in long time simulations of, e.g., chemical reaction channels
and \blk scattering setups\blk.  The approach discussed in this paper belongs to this class 2). 

The two central questions in approach 2) are, of course, how to find convenient locations for the grid-points and how to
update \blk them  in parallel with \blk the evolution of the wave function $\Psi$ given by the Schrödinger equation: 
\begin{align}
    i\partial_t\Psi_t  (X) \blk = \hat H\Psi_t(X),
    \qquad \hat H=-\half \Delta + V( \hat X \blk ),
    \label{eq:schroedinger}
\end{align}
for $t\in\mathbb R$  and $X\in \mathbb R^{Nd}$, where $\Delta$ denotes the Laplace operator with respect to the
configuration $X$, $V$ is a classical potential,  $\hat X$ is the position operator, and we use units $t\mapsto t\hbar$
and $X\mapsto (\hbar/m^{1/2})\blk X$ (for simplicity all particles are taken to have the same mass $m$).

One method is to distribute, say $M$, grid-points $Q^{(i)}_{t=0}\in\mathbb R^{Nd}$, $i=1,\ldots,M$, at initial time
$t=0$ according to the $|\Psi_{t=0}|^2$-distribution.   Thus,   regions with larger contributions to the $L^2$-norm are
sampled with higher resolution, while regions with smaller contributions are covered only by a few grid points.  In
order to ensure that the grid points follow the $|\Psi_t|^2$-distribution (a feature usually referred to as equivariance
in the context of Bohmian mechanics) \cite{detlef} one must transport them along the flux lines of the quantum
probability current~\cite{deckert}, i.e., along Bohmian trajectories $Q^{(i)}_t$ which obey the Bohmian law of
motion~\cite{detlef}
\begin{align}
    \frac{dQ^{(i)}_t}{dt}
    =
    \Im \frac{\Psi_t^*(X)\cdot\nabla\Psi_t(X)}{\Psi_t^*(X)\cdot\Psi_t(X)}\bigg|_{X=Q^{(i)}_t}.
    \label{eq:bohm}
\end{align}
Hence, Bohm's velocity law \eqref{eq:bohm} needs to be integrated simultaneously with the Schrödinger equation
\eqref{eq:schroedinger} on comoving coordinates $Q^{(i)}_t$, $i=1,\ldots,M$. Using the decomposition
$\Psi_t=\sqrt{P_t}e^{iS_t}$, the corresponding coupled set of equations \eqref{eq:schroedinger}-\eqref{eq:bohm} in
comoving coordinates takes the form \cite{wyatt}: 
\begin{align}
    \frac{d}{dt} P_t(Q^{(i)}_t) &= -P_t(Q^{(i)}_t)\Delta S_t(Q^{(i)}_t)\label{eq:prob}\\
    \frac{d}{dt} S_t(Q^{(i)}_t) &= \half \left( \frac{d}{dt} Q^{(i)}_t\right)^2 - V(Q^{(i)}_t)
    - U_t(Q^{(i)}_t)\label{eq:phase}\\
    \frac{d^2}{dt^2} Q^{(i)}_t  &= -\nabla V(Q^{(i)}_t) -\nabla
        U_t(Q^{(i)}_t)
    \label{eq:bohm2}
\end{align}
for initial value constraint
\begin{align}
    \frac{d}{dt}Q^{(i)}_t|_{t=0}=\nabla S_t(Q^{(i)}_t)|_{t=0}
    \label{eq:constraint}
\end{align}
and quantum potential
\begin{align}
    U_t(X) = -\half \frac{\Delta P_t(X)^{1/2}}{P_t(X)^{1/2}}.
    \label{eq:quantum-pot}
\end{align}
Note that constraint \eqref{eq:constraint} \blk together with \eqref{eq:bohm2} is equivalent to \eqref{eq:bohm}, in view
of the corresponding initial value problem, \blk while thanks to \eqref{eq:quantum-pot},
\eqref{eq:prob}-\eqref{eq:phase} are equivalent to \eqref{eq:schroedinger}. Numerical analysis of quantum systems with
the help of trajectories has been studied in great depth and we refer the reader to the literature, e.g.,
\cite{wyatt,chatteraj,artes,sanz}.

A further simplification can be attained when not only the grid points are distributed randomly according to $|\Psi|^2$
but when the $|\Psi|^2$-distribution itself can be approximately retrieved from the empirical distribution of the grid
point locations $Q^{(i)}_t$, via some map $P(X;Q^{(1)}_t,\ldots,Q^{(M)}_t)$ such that for all $t, X$ the approximation
\begin{align}
    P_t(X) \approx P(X;{\cal  Q}_t), \qquad {\cal  Q}_t :=
    (Q^{(1)}_t,\ldots,Q^{(M)}_t)
    \label{eq:approx}
\end{align}
holds in a suitable sense as $M\to\infty$. 
In view of the weak law of large numbers, one may think of $P(X;{\cal  Q}_t)$ as a smooth version of the empirical
distribution 
\begin{align}
     P(X;{\cal  Q}_t) \approx
     \frac1M\sum_{i=1}^M\delta^{Nd}(X-Q^{(i)}_t) ,
\end{align}
as the $Q^{(i)}_t$, $1=1,\ldots, M$, stay approximately $|\Psi_t|^2$ distributed thanks to equivariance\blk.  Once a
good candidate for $P(X;{\cal  Q}_t)$ and its derivatives is  identified, \blk equations
\eqref{eq:prob}-\eqref{eq:constraint} can be replaced by a closed set of equations for the trajectories $Q^{(i)}_t$. It
suffices, for example, to only solve the system of equations
\begin{align}
    \frac{d^2}{dt^2} Q^{(i)}_t  &= -\nabla \left[ V(X) +
        U(X; {\cal  Q}_t)
    \right]\big|_{X=Q^{(i)}_t}\label{eq:miw}
\end{align}
 under \blk the  initial \blk constraint \eqref{eq:constraint}, where now the density $P_t(X)$ in
the quantum potential \eqref{eq:quantum-pot}  is \blk  replaced by
$P(X;{\cal  Q}_t)$ so that the approximate quantum potential
reads
\begin{align}
    \label{eq:approx-potential}
    U(X;{\cal  Q}_t) 
    =
    -\half \frac{\Delta P(X;{\cal  Q}_t)^{1/2}}{P(X;{\cal  Q}_t)^{1/2}}.
\end{align}
Quantum expectation values of  an \blk
observable $f(\hat X)$ 
can now be
recovered simply \blk from the trajectories $Q^{(i)}_t$ by
\begin{align}
    \langle \Psi_t, f(\hat X) \Psi_t\rangle 
    &= \int d^{Nd}X \, P_t(X) f(X) \nn\\
    &\approx\int d^{Nd}X \, P(X;{\cal  Q}_t) f(X) \nn\\
    &\approx M^{-1} \sum_{i=1}^M f(Q^{(i)}_t).
    \label{eq:stats}
\end{align}
Several discrete and continuous versions of this approach have been proposed in the
literature~\cite{holland,poirier,parlant,schiff,prx,GhaHalWis18, sebens,bostrom,smolin,roser}.  Due the possible interpretation of
$Q^{(i)}_t$, $i=1,\ldots,M$, as $M$ coexisting ``worlds'' we follow Ref.~\cite{prx} in referring \blk to this approach
as the Many-Interacting-Worlds (MIW) \blk approach. 
Continuous versions of this
general idea~\cite{holland,poirier,parlant,schiff} predate our discrete MIW approach; see
Refs.~\cite{sebens,bostrom,smolin,roser} for continuing interest.\blk 

The MIW approach stands   or   falls   according to the possibility of   finding a good candidate $P(X;{\cal  Q}_t)$ and
the ability to maintain the quality of the approximation \eqref{eq:approx} over time for $M$ not too large. In
Ref.~\cite{prx} \blk we have presented a surprisingly simple toy model for $N=1$ particle in $d=1$ dimension 
with the ansatz
\begin{align}
    &P(Q^{(i)}; {\cal  Q}_t) \nn \\
    &:=
    \frac{1}{2}\left(\frac{1}{N(Q^{(i)}-Q^{(i-1)})}+\frac{1}{N(Q^{(i+1)}-Q^{(i)})}\right)
    \label{eq:toy-prob}
\end{align}
for $i=1,\ldots,M$, with the  ordering \blk $Q^{(i)}_0<Q^{(i+1)}_0$ 
(setting $Q^{(0)}_t=-\infty, Q^{(M+1)}_t=   +\infty$).  The above   ordering \blk is \blk 
preserved over time
because the system \eqref{eq:schroedinger} and \eqref{eq:bohm2} has a
well-defined initial value problem~\cite{BerZan05}, and hence, configuration
space trajectories cannot cross. However, instead of approximating $U_t(X)$ directly via a smoothed $P(X;{\cal  Q}_t)$, as in Eq.~(\ref{eq:approx-potential}), the method in Ref.~\cite{prx} \blk 
approximates its average (proportional to the Fisher information of $P_t(X)$), via
\begin{align} \label{eq:fisher}
\overline U_t& =\int d^{Nd}dX\,P_t(X) U_t(X) \nn \\
&= \frac{1}{8}\int P_t(X)\left|\frac{\nabla P_t(X)}{P_t(X)}\right|^2 \nn \\
&\approx \frac{1}{8}\sum_{i=1}^M \left| \frac{\nabla P(Q^{(i)}; {\cal  Q}_t)}{P(Q^{(i)}; {\cal  Q}_t)}\right|^2 .
\end{align}
Using Eq.~(\ref{eq:toy-prob}) and the corresponding discrete approximation of $\nabla P_t(X)$ for $N=d=1$, this leads to the replacement of Eq.~(\ref{eq:miw}) by the very similar form
\begin{align} \label{eq:umiw}
\frac{d^2}{dt^2} Q^{(i)}_t  &= -\nabla_{Q^{(i)}_t} \left[ V(Q^{(i)}_t) +
        U^{\textrm{MIW}}({\cal  Q}_t) 
    \right],
\end{align}
where 
\begin{align}
    &U^{\textrm{MIW}}({\cal  Q}_t) = \nn \\
    &\frac{1}{8}\sum_{i=1}^M
    \left(\frac{1}{Q^{(i+1)}-Q^{(i)}}-\frac{1}{Q^{(i)}-Q^{(i-1)}}\right)^2.
    \label{eq:miw-potential}
\end{align}
The resulting model defined by Eqs.~\eqref{eq:umiw} and \eqref{eq:miw-potential} has the nice property of conserving the total energy~\cite{prx}
\begin{align}
E=\sum_{i=1}^M \left[\half \left(\frac{dQ^{(i)}_t}{dt}\right)^2 +V(Q^{(i)}_t)\right] + U^{\textrm{MIW}}({\cal  Q}_t),
\label{eq:energy}
\end{align} 
and will be referred to as   the  1d   MIW model throughout this work. 

In Ref.~\cite{prx}, the 1d MIW model was shown to exhibit typical quantum behavior such as superposition and tunneling.
In particular, numerical implementation of the model, with a very modest number of worlds, gave good qualitative
agreement in the case of the time-evolution of two superposed Gaussians   (representing double-slit interference).  In
addition, numerical testing showed good quantitative agreement for the computation of ground states, and convergence in
the limit $M\to\infty$ has been \blk proven for a harmonic potential~\cite{mckeague}. \blk

The goal of this paper is to develop the MIW approach further, to treat more than one degree of freedom.
Section~\ref{sec:generalization} \blk provides a general model for any finite   number of   degrees of freedom, i.e.,
finite  particle numbers and spatial dimensions. In the spirit of this general approach we then present \blk numerical
algorithms for finding energy eigenstates. \blk First, we consider the 1d case in section~\ref{sec:1d}, to benchmark our
new method against the original 1d MIW model, using the harmonic and the Pöschl-Teller potentials, and we present
results for both ground and excited states. \blk Second, we 
generalize our method of finding ground states to $d=2$ dimensions in section~\ref{sec:2d}. In particular, we discuss
our numerical results again for the harmonic and Pöschl-Teller potentials. In comparison  to the harmonic potential, the
Pöschl-Teller potential is only weakly confining, which makes the lack of information at spatial infinity much more
prominent in numerical simulations. We discuss how this problem can be successfully addressed in our approach.

The numerical methods and simulations reported here are based on results in~\cite{hannes}. Independent calculations have
been made very recently by Sturniolo~\cite{sturn}, for the ground states of higher dimensional systems in the framework
of the MIW approach, which we comment on briefly in section~\ref{sec:conclusion}.

\section{Generalization to arbitrarily many degrees of freedom}
\label{sec:generalization}


A formal extension of the 1d MIW model, to a system of $N$ particles moving in
$d$ spatial dimensions, is given by retaining  the equations of motion,
Eq.~(\ref{eq:umiw}), but \blk generalizing Eq.~(\ref{eq:miw-potential}) to
\begin{align} \label{eq:genmiw}
 U^{\textrm{MIW}}({\cal  Q}_t) 
    = \frac{1}{8}\sum_{i=1}^M \left| \frac{\nabla P(Q^{(i)}_t; {\cal  Q}_t)}{P(Q^{(i)}; {\cal  Q}_t)}\right|^2  ,
\end{align}
for suitable approximations $P(Q^{(i)}_t; {\cal  Q}_t)$ and \blk $\nabla P(Q^{(i)}_t; {\cal  Q}_t)$, of $P_t(Q^{(i)}_t)$
and its derivative, respectively~\cite{prx}. \blk We now show how to construct these two approximations in turn. \blk 

\subsection{Approximating the probability density} \label{sec:papprox}

We will consider two related approaches here, based on triangulation and cells, respectively.

\subsubsection{Triangulation method}

\begin{figure*} 
    \begin{center}
        \includegraphics{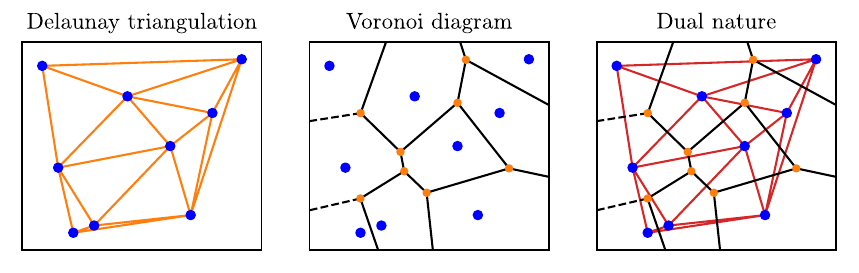}
    \caption[TrianglesAndCells]{
        \label{fig:voronoi}
        The left hand panel shows a Delaunay triangulation of configuration space
        for a set of worlds (blue circles) at a given time. Each triangle (orange
        edges) is chosen in such a way that no worlds lie inside the circum-sphere
        of any triangle.  The middle panel shows its dual graph, corresponding to
        partitioning configuration space into Voronoi cells (see definition
        \eqref{eq:voronoi} below). The corresponding cell boundaries are formed
        by hyperplanes (black lines) which bisect the triangulation lines. The
        right panel illustrates the duality of these graphs.
    }
    \end{center}
\end{figure*}

 The worlds or trajectories $Q^{(i)}_t$ lie in the $D$-dimensional
 configuration space $\mathbb R^{Nd}$ with $D:=Nd$. This configuration space
 can  be \blk partitioned into a network of $D$-tetrahedra having the worlds as
 vertices, together with a single exterior region. For $D=2$ this corresponds
 to a triangulation of the configuration space, together with an exterior
 region. Such a triangulation is depicted in the left hand panel  of
 figure~\ref{fig:voronoi} (purple lines), corresponding to a Delaunay
 triangulation~\cite{delaunay}. Efficient algorithms are known for establishing
 such triangulations~\cite{delaunay}.

For a given triangulation, let $\{T_{i,j}\}$ denote the set of $D$-tetrahedra
(`triangles') sharing $Q^{(i)}$ as a common vertex, at a given time $t$.  Here
we have dropped the explicit time label on $Q^{(i)}$ \blk for convenience.
Now, for  a \blk sufficiently smooth function $f(X)$ on configuration space one can,
similarly to Eq.~(\ref{eq:stats}), \blk approximate  its \blk average 
via
\begin{align}
\frac{1}{M}\sum_{i=1}^M {f}(Q^{(i)}) &\approx \int dX\,P(X)\,{f}(X) \nn \\
&=\frac{1}{D+1} \nn \sum_{i,j}  \int_{T_{i,j}} dX\,P(X)\,{f}(X)\\
&\approx \frac{1}{D+1}\sum_{i,j} |T_{i,j}|\,P(Q^{(i)})\,{f}(Q^{(i)}) ,
\end{align}
where $|T_{i,j}|$ denotes the volume of the tetradron $T_{i,j}$, and the factor of $D+1$ arises because every tetrahedron is counted once for each of its $D+1$ vertices.  Hence, equating these expressions for arbitrary $f(X)$, a suitable approximation for the probability density at $Q^{(i)}$ is given by
\begin{equation} \label{eq:triangle}
P_{\rm tri}(Q^{(i)};{\cal  Q})  = \blk \frac{D+1}{M \sum_j |T_{i,j}|} .
\end{equation}
This reduces to Eq.~(\ref{eq:toy-prob}) for the 1d MIW model when $D=Nd=1$.

\subsubsection{Cell method}
\label{sec:cell}

An alternative to placing worlds at the vertices of a triangulation is to instead place each world within an individual cell, where the cells partition the configuration space.  For example, for a Delaunay triangulation such cells can be chosen as the dual graph, corresponding to Voronoi cells~\cite{delaunay}. An example is depicted in the right panel of figure~\ref{fig:voronoi}. Note that some cells, corresponding to worlds at the edges, are infinite in extent.

Such a partitioning leads to the alternative approximation
\beq
P_{\rm cell}(Q^{(i)};{\cal  Q}) = \frac{1}{M |\cell_i|} 
\label{eq:cell}
\eeq
for the probability density at $Q^{(i)}$, where $|\cell_i|$ denotes the configuration space volume of the cell containing trajectory $Q^{(i)}$. 

\subsection{Approximating the derivative of the probability density}

Equation~(\ref{eq:genmiw}) further requires finding a suitable approximation for the derivative $\nabla P(Q^{(i)})$ at each time (where again we suppress the explicit dependence on time for convenience). This derivative has $D=Nd$ independent components, and hence we need to consider, for each trajectory $Q^{(i)}$ at that time, the change in probability in at least $D$ different directions. These directions could be chosen, for example, to be those which join $Q^{(i)}$ to its $D$ closest neighbours, independently of the method used to estimate the density itself (e.g., via triangulation or cells).   An alternative choice is to use the directions corresponding to all (or some) edges of a given triangulation which have $Q^{(i)}$ as a vertex. Yet another choice is to use the directions corresponding to all (or some) worlds that share a cell boundary with $Q^{(i)}$.  

Here we will be quite general, and only suppose that $C_i\geq D$ neighbouring
 configurations \blk 
or worlds are used
to estimate $\nabla P(Q^{(i)})$, where these may be selected by any of the
means above. Let $\{Q^{(i,s)}\}$ denote these $C_i$  configurations, \blk and define
the corresponding set of vectors 
\beq
v^{(i,s)}:=Q^{(i,s)}-Q^{(i)} .
\eeq 
By construction, these form a (typically overcomplete) basis set in configuration space.  Now, writing 
\beq 
v^{(i,s)}=\sum_{k=1}^D A_{sk}e_k
\eeq
relative to some orthonormal basis set $\{e_k\}$,  one has a corresponding set of dual basis vectors $\tilde v^{(i,s)}:=\sum_k {\tilde A}_{sk} e_k$, with $\tilde A:=A(A^\top A)^{-1}$. This dual basis satisfies the completeness property
\beq \sum_s \tilde v^{(i,s)}\,(v^{(i,s)})^\top = I_D ,\eeq
where  $\top$ denotes the transpose and $I_D$ is the $D\times D$ identity matrix. 
Using $P(Q^{(i,s)})-P(Q^{(i)})\approx v^{(i,s)}\cdot\nabla P(Q^{(i)})$, it follows immediately that one has 
\begin{align} 
\nabla P(Q^{(i)}) &=  \sum_s \tilde v^{(i,s)} \,(v^{(i,s)})^\top \,\nabla P(Q^{(i)}) \nn \\
&\approx \sum_s  \left[ P(Q^{(i,s)})-P(Q^{(i)}) \right]\,\tilde v^{(i,s)} .
\label{tetgrad}
\end{align}
For any suitable approximation of $P(Q^{(i)};Q)$, such as in Eqs.~(\ref{eq:triangle}) or~(\ref{eq:cell}), one then has the corresponding approximation
\begin{align} \label{eq:derivative}
 \nabla P(Q^{(i)};{\cal  Q}) = \sum_s  \left[ P(Q^{(i,s)};{\cal  Q})-P(Q^{(i)};{\cal  Q}) \right]\,\tilde v^{(i,s)}
\end{align}
of the derivative. This may be inserted into Eq.~(\ref{eq:genmiw}) to obtain the corresponding MIW potential function $U^{\textrm{MIW}}({\cal  Q}_t)$ at any given time $t$.

%
%

\section{A numerical implementation for ground states}
\label{sec:numerics}

Following the spirit of the general approach in
section~\ref{sec:generalization} we will now provide numerical implementations
in 1d and 2d which are based on the work in Ref.~\cite{hannes}. \blk %
We will concentrate primarily on the numerical determination of ground state energies and  distributions, although we do also consider excited states in the 1d case. \blk 

While the approach given in section~\ref{sec:generalization} \blk is precise and general, we encountered several
problems in its direct numerical implementation.  For example, the construction of the dual basis set $\{\tilde
v^{(i,s)}\}$ appearing in Eq.~(\ref{eq:derivative}) requires computation of the inverse of the $C_i\times C_i$ matrix
$(A^\top A)^{-1}$  for each of the $M$ worlds. Recalling that $C_i\geq D=Nd$, this alone requires $O(MN^3d^3)$
calculations at each time step. \blk 
Moreover, unlike the one-dimensional case, the set of neighboring \blk configurations or worlds used to define
triangulations, partitionings, derivatives, etc., can change over time, and thus requires constant updating.  
 
These problems originate from the use of neighboring worlds for obtaining approximations of $P_t(X)$ and its derivative
at a given world $Q^{(i)}_t$. The selection of finitely many nearest neighbors inevitably provokes discontinuous changes
throughout the dynamics. Though for very large $M$ one may expect that these sudden jumps may have only have little
impact on the overall dynamics of worlds, this is not the case for lower \blk values of $M$. There, \blk these small
discontinuities may cause oscillations in the world configurations $Q^{(i)}_t$ which are not damped, and which propagate
through the whole system until the numerical simulation breaks down. This forced us to replace this discrete notion of
nearest neighbors, in computing the approximate $P_t(X)$ and its derivatives, by something more smooth. 
Enforcing some sort of smoothness may also come as no surprise: Even if
the grid points may sample well regions in which $P_t(X)$, i.e., $|\Psi_t|^2$, is large, and thus, potentially increase
the precision in the $L^2$-norm sense without the need of too many samples,  the required precision in the pointwise
sense in~\eqref{eq:miw}, i.e., \eqref{eq:bohm} cannot be guaranteed, unless some prior knowledge on the smoothness is
available. \blk 

It turns out that finding a smooth distribution that approximates the empirical distribution is an old problem,
discussed thoroughly \blk in the classical literature; see~\cite{izenman,scott,abramson,elgmmal} for an overview. One
general and, for many settings,  very robust technique is  so-called smooth kernel density estimation, which we
introduce first.

The density estimator for a given distribution of worlds
${\cal  Q}=(Q^{(1)},\ldots, Q^{(M)})$ is given by a sum of the form
\begin{align}
    P_h(X;{\cal  Q}) := \frac{1}{\widetilde M} \sum_{i=1}^{\widetilde M} \frac{1}{h_i}
    K\left(\frac{X-\widetilde Q^{(i)}}{h_i}\right).
    \label{eq:estimator}
\end{align}
Here $\{\widetilde Q^{(i)}\}$ is a set of $\tilde M$ points in configuration
space determined by ${\cal  Q}$; the $h_i$ are  width parameters (usually referred
to as bandwidths) \blk similarly determined
by ${\cal  Q}$; and $K$ is a smooth kernel function that fulfills $\int d^{Nd}X
\,K(X)=1$.  Note that $P_h(X;{\cal  Q})$ is automatically normalized.  Although, this leaves a lot of freedom, in this work we will only
focus on the Gaussian kernel $K(X)=(2\pi)^{-{Nd/2}}\exp(-\half X\tp X)$, 
for
which $\widetilde Q^{(i)}$ takes the role of a mean and $h_i^2  I_{Nd} \blk$ defines a corresponding covariance matrix
(we do not explore \blk more general covariance matrices here). Considering that the Schrödinger propagator is given by
a Gaussian~\cite{detlef}, this seems like a canonical choice. 

The idea behind ansatz \eqref{eq:estimator} is to allow for varying widths $h_i$, well-adapted to regions of high and
low empirical density in the vicinity of conveniently chosen locations $\widetilde Q^{(i)}$. If the empirical density is
low in the neighborhood of $\widetilde Q^{(i)}$, one chooses large values of $h_i$, i.e., broad kernel functions, and if
the density is high, one chooses small values, i.e., peaked kernel functions. We will come back to the question of
choosing optimal $\widetilde Q^{(i)}$ and $h_i$ later, in sections~\ref{sec:1d} and \ref{sec:2d}. We first show how
density estimation may be used in an algorithm for calculating ground state properties.

\subsection{Gaussian kernel algorithm}
\label{sec:kernel}

Once the choice for the $\widetilde Q^{(i)}$ and $h_i$ is settled, an algorithm
for finding ground states can be given in terms of the following iteration: 
\begin{enumerate}
    \item Start with any initial distribution of $M$ worlds
        ${\cal  Q}_0=\{Q^{(1)}_0,\ldots,Q^{(M)}_0\}$ and choose a suitably small \blk time step $\Delta t>0$.
    \item From ${\cal  Q}_0$, compute the approximate potential
        \eqref{eq:approx-potential} in which the approximate density
        $P(X;{\cal  Q}_t)$ is replaced by $P_h(X;{\cal  Q}_0)$ given in \eqref{eq:estimator}.
    \item Integrate the second order equation of motion \eqref{eq:miw} up to
        time $\Delta t$ with zero initial velocities $\dot Q^{(i)}_0=0$, to obtain
        a new empirical distribution  ${\cal  Q}_{\Delta t}$.
    \item Replace ${\cal  Q}_0$ by ${\cal  Q}_{\Delta t}$ and go back to step 2 until a
        predefined stopping condition is met (e.g. given by an appropriate
        measure of convergence).\blk
\end{enumerate}
We shall refer to this algorithm as the {\it Gaussian kernel algorithm}. The numerical implementation used in this work is provided in \cite{git}.

A similar algorithm was discussed for the 1d MIW model in Ref.~\cite{prx}. \blk The reason why convergence can be expected is that
in every integration step of \eqref{eq:miw} the initial velocities are set to
zero.  This introduces a loss of energy, as after each 
integration step 3 above the total energy 
\begin{align}
    &E_{\rm  kin}(\Delta t) + E_{\rm pot}(\Delta t)
    = \sum_{i=1}^M 
    \bigg[\half \left(\dot Q_{\Delta t}^{(i)}\right)^2 \nn \\
    &- \int_0^{\Delta t} ds \,
    \dot Q^{(i)}_s \cdot \nabla [V(X)+U(X;{\cal  Q}_s)]_{X=Q^{(i)}_s}\bigg]
    \label{eq:energysum}
\end{align}
is reduced by the positive kinetic energy $E_{\rm kin}$. Hence, during the
iteration of the algorithm the configuration of worlds ${\cal  Q}$ will arrange itself
to find a local minimum of $E_{\rm  pot}(\Delta t)$. Providing that the potential $V(X)$ is
confining, e.g., as in the case of a harmonic potential, it will work to focus
the worlds, while the potential $U(X;{\cal  Q})$ will work against clustering of worlds (cf.~ Ref.\blk~\cite{prx}).
Since the integration time step $\Delta t$ is small, and near a local minimum
the velocities $\dot Q^{(i)}_s$ in Eq.~(\ref{eq:energysum}) can also be expected to be
small, a local minimum of $E_{\rm  pot}$ should then fulfill
\begin{align}
    \label{eq:equilibrium-cond}
    \nabla [V(X)+U(X;{\cal  Q})]_{X=Q^{(i)}} \approx 0 ,
\end{align}
which according to the Bohmian equation of motion  corresponds to a stationary state~\cite{detlef}. If $V(X)$ has only
one local minimum one can, therefore, expect that the algorithm converges to a configuration of worlds ${\cal  Q}$ that
is distributed according to $|\Psi|^2$, where  $\Psi$ is the ground state of the system with Hamiltonian $\hat H$ as per
\eqref{eq:schroedinger}.

The main difference between the above algorithm, employing Gaussian kernels, and the MIW algorithm given in
Ref.~\cite{prx}, \blk is that the latter does not use a density estimator but instead computes  forces as per
\eqref{eq:umiw}, using the MIW potential \eqref{eq:miw-potential}, where the latter is conservative as per
Eq.~(\ref{eq:energy}). One of the advantages of the Gaussian kernel model introduced here is that its form readily
generalizes to any number of degrees of freedom $Nd$ without sacrificing smoothness. In contrast, the form of
\eqref{eq:umiw} and its generalization via  Eq.\blk ~(\ref{eq:genmiw}) depend on the use discrete derivatives, defined
via finitely many neighbouring worlds, which leads to  continuity issues as discussed at the beginning of this section.

\subsection{Application to 1d ground states}
\label{sec:1d}

The goal of this section is to provide a numerical implementation of the
Gaussian kernel algorithm for $Nd=1$. This  allows a comparison with the MIW
algorithm studied in  Ref.~\cite{prx}\blk, and will provide the basis for the generalization
to $Nd=2$ in section~\ref{sec:2d}.  

The first question we must address is the choice of the $\widetilde Q^{(i)}$ and $h_i$ in Eq.~(\ref{eq:estimator}), on which the performance of the algorithm will crucially
depend.  Several explicit forms for $\widetilde Q^{(i)}$
and $h_i$ have been considered in the literature on kernel density estimators, for various situations.
For example,  Ref.~\blk~\cite{scott} discusses an explicit dependence of $h_i$ on the
$k$-th nearest neighbors distance w.r.t.\ $\widetilde Q^{(i)}=Q^{(i)}$.
However, for reasons discussed above, we want to avoid a dependence on
discontinuous quantities such as the nearest neighbor distance, as much as
possible. Hence we propose another approach here.

For the case $Nd=1$ (although not for higher values), it is important to observe that the world configurations
${\cal  Q}_{\Delta t}^{(i)}$ found in step 3 of the Gaussian kernel algorithm in Sec.~\ref{sec:kernel} should be
good approximations to Bohmian trajectories, and as such may not cross~\cite{deckert}.  In particular, if a crossing occurs then the trajectories ${\cal  Q}_{\Delta t}^{(i)}$ are no longer trustworthy, and there is no reason why in future iterations of the
algorithm they will converge to a sensible distribution.  However, for very large $M$, crossings may easily occur due to numerical errors. Hence, for $Nd=1$ it is important
to implement a mechanism that effectively works against such
catastrophic crossing events.  We note that \blk such events are  much more
suppressed in the 1d MIW model, due to the singular \blk repulsion between neighbouring
worlds~\cite{prx}.

One mechanism that we found to work well is
implemented by choosing the $\widetilde Q^{(i)}$ to be located midway between the actual world
configurations $Q^{(i)}$:
\begin{align} \label{eq:1dmeans}
    \widetilde Q^{(i)}:=\half(Q^{(i)}+Q^{(i+1)}),
\end{align}
for $i=1,\ldots,\widetilde M=M-1$. Provided the width parameters $h_i$ are set
appropriately, this choice induces a peak in the  density $P_h(X;{\cal  Q})$ to build up between any two
approaching worlds $Q^{(i)}_{\Delta t}$ and $Q^{(i+1)}_{\Delta t}$. By virtue
of the second derivative \eqref{eq:approx-potential} the corresponding quantum
force on the right-hand side of \eqref{eq:miw} then acts to repel these two
approaching worlds; {\it cf.}~Ref.~\cite{deckert}.

\begin{figure*}
	\begin{center}
		\includegraphics{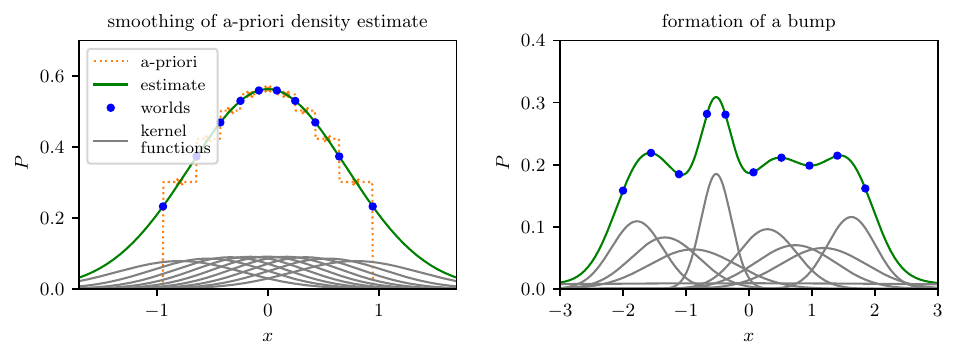}
		\caption[Gaussian Kernel Estimator]{
			\label{fig:gaussinterpolation}
			The left panel illustrates 1d kernel
			density estimation, as per Eqs.~(\ref{eq:estimator}), (\ref{eq:1dmeans}) and~(\ref{eq:P-constraint}), for the case of a Gaussian distributed
			configuration of worlds ${\cal  Q}$. The $M=10$ worlds $Q^{(i)}$ are shown as
			red dots, the $\tilde M=9$ midpoints
			$\tilde Q^{(i)}$ in Eq.~(\ref{eq:1dmeans}) correspond to the blue
			crosses, and the {\it a priori} density estimates $p_i$ are plotted as
			the blue dotted line. The corresponding Gaussian kernels are shown in
			gray, and sum to  $P_h(X;{\cal  Q})$ (depicted by the green curve) as per
			Eq.~(\ref{eq:estimator}). The right panel shows how the density
			estimate naturally develops a bump between two nearby
			worlds (see also Sec.~\ref{sec:1d}).  }
	\end{center}
\end{figure*}

In order to determine good 
bandwidths $h_i$ we  constrain \blk the estimator
ansatz \eqref{eq:estimator} by
\begin{align}
    \label{eq:P-constraint}
    P_h(\widetilde Q^{(i)}; {\cal  Q}) \overset{!}{=} \frac{1}{M+1}\frac{1}{Q^{(i+1)}-Q^{(i)}} =: p_i
\end{align}
for all $i=1,\ldots, M-1$. The {a priori} estimate of the density $p_i$ between
the two worlds is, of course, very much related to the {a priori} density
that was used to construct the MIW model;
{\it cf.}~\eqref{eq:toy-prob}.  This
formula stems from the mean value theorem and can be seen as special case of
\eqref{eq:cell}. \blk Instead
of trying to infer an analytic solution for this constraint (which  may not
be \blk possible in general), we
implement the  recursion relation 
\begin{align}
    h_i \leftarrow h_i \frac{P_h(\widetilde Q^{(i)};{\cal  Q})}{p_i},
    \label{eq:bandwidths}
\end{align}
which in typical situations considered in this paper gives good
results after just a few iterations. Figure~\ref{fig:gaussinterpolation}
illustrates the density estimation in two cases (see the figure caption for further discussion).

\begin{figure*}
    \begin{center}
        \includegraphics{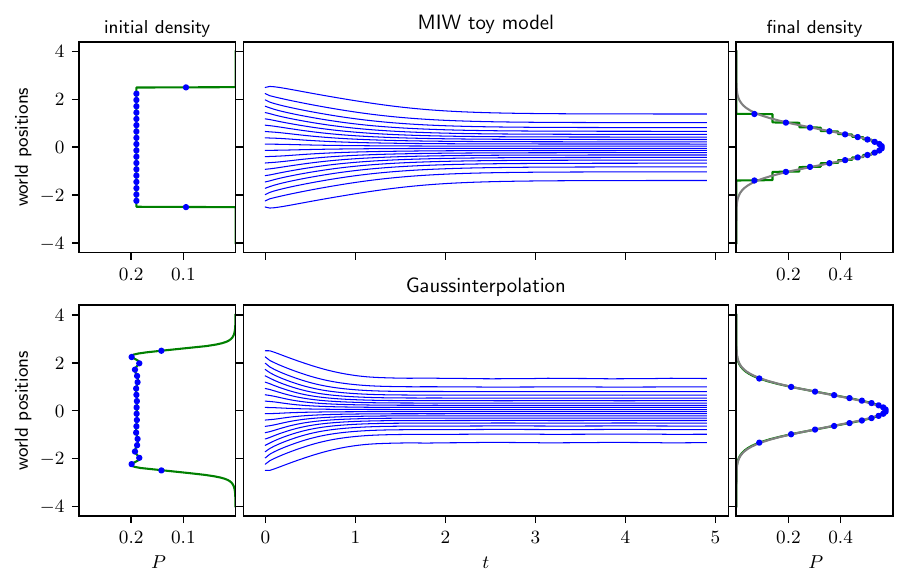}
        \caption{
            \label{fig:evolution} Comparison of the convergence of the Gaussian
            kernel algorithm (top panel) and the MIW algorithm (bottom panel)
            in the case of a 1d harmonic potential $V(X)=\half \hbar^2\omega^2
            X^2$, $M=20$ worlds, and $\Delta t = 4.9\cdot 10^{-5}$ in units of
            $\hbar$ over $10^5$ iterations  (see also Sec.~\ref{sec:1d}). \blk
            The two plots on the left illustrate the initial spacing of the
            worlds ${\cal Q}$, which was chosen to be uniform (blue dots), and
            the density estimate using \eqref{eq:estimator} and
            \eqref{eq:bandwidths} respectively (green curves). The two middle
            plots show the evolving configuration of the worlds during the
            iteration of the algorithms and the plots on the right show the
            final densities after the convergence. 
        }
    \end{center}
\end{figure*}

\begin{figure*}
    \begin{center}
        \includegraphics{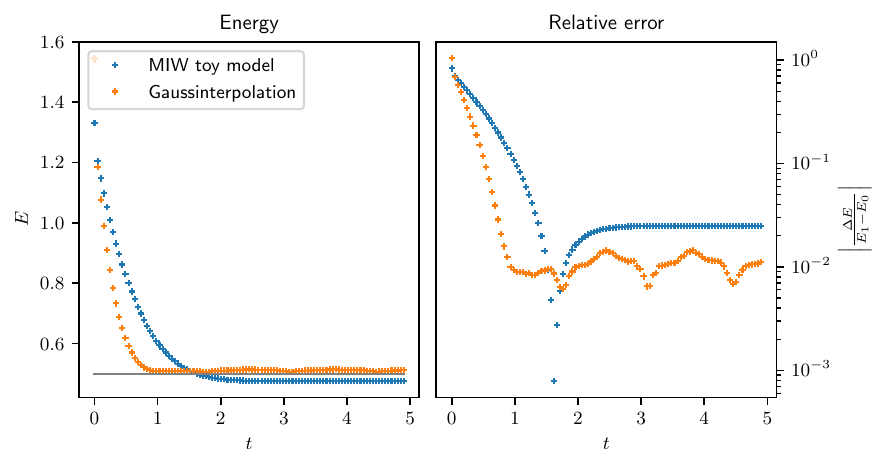}
        \caption{
    \label{fig:energy}
            Convergence of the total energy and the relative error for the
            Gaussian kernel algorithm (blue $+$ markers) and the  1d \blk MIW
            model \blk algorithm (orange $+$ markers), for the same setting as
            in figure~\ref{fig:evolution} (see also Sec.~\ref{sec:1d}). The
            energy is plotted in units of $\hbar\omega$.  The relative
            error is given as difference w.r.t. the true ground state and
            normalized w.r.t.\ the energy difference between the first excited
            and ground state energy, i.e., $E_1=(3/2)\blk\hbar\omega$ and
            $E_0=(1/2)\blk\hbar\omega$, respectively. 
        }
    \end{center}
\end{figure*}

For the numerical implementation of the Gaussian kernel algorithm we chose a
similar setting as for the MIW algorithm discussed in  Ref.~\cite{prx}\blk:
$V(X)=\half \hbar^2\omega^2 X^2$ for magnitude of $\hbar^2\omega^2$ being one,
$M=20$ worlds, and a time step of $\Delta t=4.9\cdot 10^{-5}s^{-1}\,\hbar$ over
$10^5$ iterations.  
The convergence of the configuration of worlds ${\cal  Q}$ is illustrated in
the top panel of figure~\ref{fig:evolution}, with the corresponding result for
the MIW algorithm shown in the bottom panel for comparison. 
As a measure of the performance and accuracy of the algorithms we used the
convergence of ground state energy which is illustrated in
figure~\ref{fig:energy}. The  1d MIW model
algorithm was already shown~\cite{prx} \blk to converge to a ground state
energy of $\half (1-M^{-1})$, corresponding to an asymptotic relative
error of $M^{-1}=5\cdot 10^{-2}$, as depicted in the right panel of the figure. In
contrast, the Gaussian kernel algorithm gives a smaller (albeit oscillating)
relative error as can be seen in figure~\ref{fig:energy}.  Whereas the MIW algorithm
systematically underestimates the ground state energy as above, the Gaussian
kernel algorithm appears to systematically overestimate the energy.
Furthermore, the Gaussian kernel algorithm converges rather faster to the
ground state distribution. This behavior is very likely due to the fact that
the Gaussian kernel function are well adapted to the approximate the ground
state density for a harmonic potential. Overall, in \cite{hannes}, the Gaussian
kernel algorithm was tested in various scenarios (and for different potentials)
and performed quite robustly in approximating the respective ground
states. 

\subsection{Extension to 1d excited states}

We observed the above algorithm to also converge quickly for various other
potentials. Hence, a next natural question is if one can also find excited
states by a similar approach. After all, for excited states one would similarly have
to look for a distribution of worlds such that \eqref{eq:equilibrium-cond} is
fulfilled. However, we found that this does not work out-of-the-box. 
One reason is that, according to the Gaussian kernel algorithm
introduced above, the energy of the configuration of worlds is monotonically
decreasing  in integration time. \blk Any numerical inaccuracy in the configuration of worlds
approximating  the distributed of square modulus of an excited wave function
(which for $N<\infty$ is generic) may likely cause the energy to decrease below
the one of the excited states in the next integration step, after which the
energy will decrease further until a ground state configuration is approached.
To find a particular excited state one must therefore
search for a ground state w.r.t.\ a Hilbert space that excludes the span
of all eigenstates below a certain energy level. 

An intuition for how to implement such a restriction comes from
Courant's old observation \cite{courant} that (assuming a sufficiently regular potential) the
number of nodes of the wave function of the $n$-th excited state is divided in
no more than $n$ subdomains of the configuration space, and in particular, the
ground state does not admit any nodes; see~\cite{aconda} for a modern
discussion. Exploiting this observation, we may restrict the search for a
stationary with a certain predetermined number of nodes by replacing our
density estimator \eqref{eq:estimator} by a new one that imposes the nodes
manually:
\begin{align}
    & Z P_h(X;{\cal  Q}) := \sum_{i=1}^{\widetilde M} \frac{1}{h_i}
    K\left(\frac{X-\widetilde Q^{(i)}}{h_i}\right) \nn \\
    & - \sum_{i=1}^{\bar M} \frac{1}{\bar h_i}
    K\left(\frac{X-\bar Q^{(i)}}{\bar h_i}\right). 
    \label{eq:estimator-nodes}
\end{align}
Here, $\bar M$ denotes the number of enforced nodes and $\bar Q^{(i)}$ their
positions.  Note the minus sign in front of the second summand and the
normalization constant $Z$ on the left-hand side. The additional parameters
$\bar h_i$ play the same role as the $h_i$ and determine the behavior of the
world distribution near a node. Unfortunately, this new density estimator may
assume negative values and therefore may in general not give rise to a proper
probability density. However, in our first trials with the harmonic and
Pöschl-Teller potential this fact has turned out to be negligible for the
performance of the algorithm. 

In 1d using the Gaussian kernel algorithm as described above, it turns out to
be sufficient to work with the old density estimator \eqref{eq:estimator} 
for the
same $\widetilde M$ while allowing for negative $h_i$ (since the kernels are
symmetric).  Instead of enforcing the nodes at the level of the density
estimator as was done \blk in \eqref{eq:estimator-nodes}  they can be enforced by changing the {\it a priori} estimate
\eqref{eq:P-constraint}  into 
\begin{align}
    \label{eq:P-constraintnodes}
    P_h(\widetilde Q^{(i)}; {\cal  Q}) \overset{!}{=}
    \frac{1}{M+1}\frac{1}{Q^{(i+1)}-Q^{(i)}} \chi_i,
\end{align}
where $\chi_i=0$ if there is supposed to be a node between
 worlds \blk $\widetilde Q^{(i)}$ and $\widetilde Q^{(i+1)}$, while otherwise $\chi_i=1$. The
recursion \eqref{eq:bandwidths} must  also \blk be changed, in order to allow for vanishing
right-hand side of \eqref{eq:P-constraint}. One possible choice is given by
\begin{align}
    h_i \leftarrow \frac{K(0)}{p_i - P_h(\widetilde Q^{(i)};{\cal  Q}) + h_i^{-1}
    K(0)}.
    \label{eq:bandwidths2}
\end{align}

\begin{figure*} 
    \begin{center}
        \includegraphics{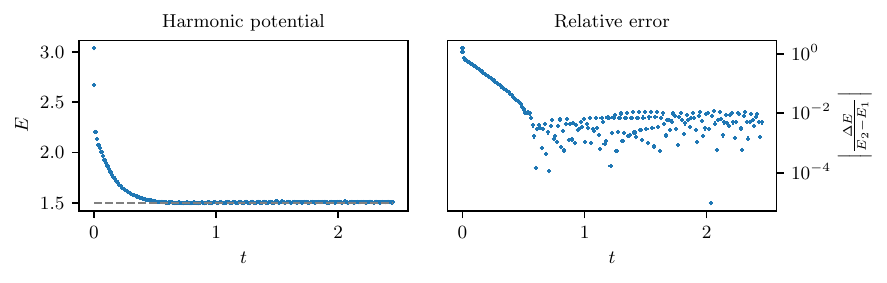}
        \includegraphics{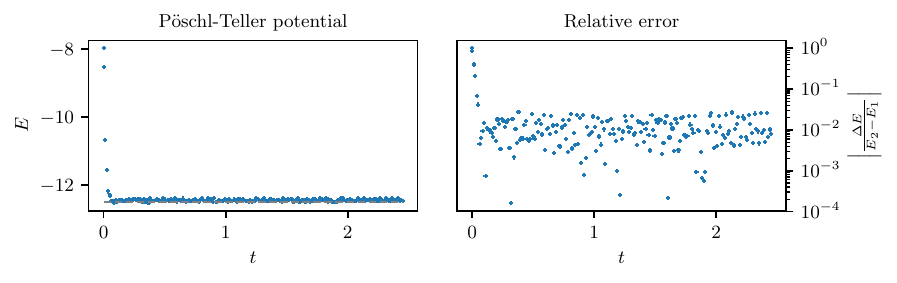}
    \caption{
    \label{fig:excited}
    The upper and lower left plots depict the convergence to the first excited
    state of the harmonic oscillator potential $V(X)=\half \hbar^2\omega^2 X^2$
    and the Pöschl-Teller potential
    $V(X)=\frac{\alpha^2}{2}\frac{\lambda(\lambda+1)}{\rm cosh^2(\alpha X)}$
    for $\lambda=6$,
    respectively. The energy in the upper plot is plotted in units of
    $\hbar\omega$ and for the lower one in units of $\alpha^2 \hbar^2 / m$. For
    both cases we used an integration time step
    $\Delta t=2.45\cdot 10^{-5}$ in units of $\hbar$ and $10^5$ iterations.
    Furthermore, the
    respective relative error is illustrated in the plots on the right and
    again given in terms difference w.r.t.\ the first excited state and
    normalized w.r.t.\ the energy difference between the first excited and
    ground state energy, i.e., $E_2$ and $E_1$. Note that for the harmonic oscillator
potential one has $E_2=(5/2)\blk\hbar\omega$ and $E_0=(3/2)\blk\hbar\omega$, and for the
Pöschl-Teller potential $E_2=-8\cdot\alpha^2 \hbar^2 / m$ and $E_1=-12.5\cdot\alpha^2 \hbar^2 / m$.}
    \end{center}
\end{figure*}

In~\cite{hannes} several choices \blk (including the above) have been
numerically studied and benchmarked. Here we  give \blk the results found for
the harmonic potential $V(X)=\half \hbar^2\omega^2 X^2$ and the Pöschl-Teller
potential $V(X)=\frac{\alpha^2}{2}\frac{\lambda(\lambda+1)}{\rm cosh^2(\alpha
X)}$, for which we chose $\lambda=6$; see figure~\ref{fig:excited}. For both potentials we have
set up the Gaussian kernel algorithm to find the first excited ground state.
This was done by exploiting the radial symmetry of the potentials which
dictates that one node has to be enforced at the origin. 
For both cases we found a fairly quick convergence up to an accuracy that is of the
same order of magnitude as the one found in the computation of the ground
states (see figure~\ref{fig:energy}). After this convergence phase the
computed energy values starts to oscillate with high frequency but small
amplitude (observed also in the ground state computation; see
figure~\ref{fig:energy}). Hence, it lies near that it is caused by the crude
iteration scheme we use to find the bandwidths $h_i$; see \eqref{eq:bandwidths}
and \eqref{eq:bandwidths2}. \blk

\subsection{Application to 2d ground states}
\label{sec:2d}

The Gaussian kernel algorithm defined in section~\ref{sec:kernel} is applicable to any number of degrees of freedom
$Nd$. However, the choices for $\widetilde Q^{(i)}$ and $h_i$ in section~\ref{sec:1d} were \blk constructed for the case
$Nd=1$, and hence need to be generalized.  

It will be recalled that for $Nd=1$ we chose $\widetilde Q^{(i)}$ according to \eqref{eq:1dmeans} because we needed a
mechanism to prevent crossing of worlds. However, for $Nd>1$ the dynamics is much less constrained---worlds can move so
as to exchange positions in configuration space---so that we can be less careful and simply choose
\begin{align}
    \widetilde Q^{(i)} = Q^{(i)}
\end{align}
for $i=1,\ldots,\widetilde M=M$. Note that this prescription is independent of the precise number of degrees of freedom $Nd$. 

Next, we need
to consider an appropriate generalization of constraint
\eqref{eq:P-constraint}, to prescribe the width parameters $h_i$ for a given 
configuration of worlds ${\cal  Q}$. This is most easily effected by finding a suitable
replacement for the {{\it a priori}} density estimate $p_i$ in
Eq.~(\ref{eq:P-constraint}). Any of the general methods in
section~\ref{sec:papprox} is convenient in this regard, and we will in
particular 
follow the method in section~\ref{sec:cell}, based on Eq.~\eqref{eq:cell}.
 In order to use
\eqref{eq:cell} we need to specify a subdivision of the configuration space into cells.
As mentioned earlier, there are various subdivision methods available, and in our numerical
implementation we have chosen the Voronoi subdivision method~\cite{delaunay}.

In particular, for the
configuration space $\mathbb R^{Nd}$ and a configuration of worlds
$Q^{(i)}\in\mathbb R^{Nd}$, the Voronoi cell containing the world $Q^{(i)}$ is
defined by
\begin{align}
    &\cell_i := \nonumber\\
    &\bigg\{
        X\in \mathbb R^{Nd} : \|X-Q^{(i)}\| < \|X-Q^{(j)}\| \,\forall j\neq i
    \bigg\}
    \label{eq:voronoi}
\end{align}
We will call $\cell_i$ an \emph{inner cell} if it is bounded and an
\emph{outer cell} if it is unbounded. By definition the Voronoi cells form a
subdivision \blk of configuration space $\mathbb R^{Nd}$: $\bigcup_i
\overline{\cell_i} = \mathbb
R^{Nd}$ and $\cell_i\cap\cell_j=\emptyset$ for $j\neq i$. As discussed in
section~\ref{sec:generalization},
it can be shown that the graph generated by the edges of all Voronoi cells is
the dual graph to the one generated by the edges of a Delaunay triangulation
(see also figure~\ref{fig:voronoi}). The
Voronoi subdivision is well adapted to our problem of finding an {{\it a priori}}
density such as \eqref{eq:cell} from an empirical distribution defined by ${\cal  Q}$
as it very naturally incorporates a measure of proximity in configuration space. In accordance
with \eqref{eq:cell}, we shall therefore use the corresponding {{\it a priori}} density constraint
\begin{align}
    \label{eq:genP-constraint}
       P_h(\widetilde Q^{(i)}; {\cal  Q}) \overset{!}{=} \frac{1}{M}\frac{1}{|\cell_i|}  =: p_i 
\end{align}
in place of Eq.~(\ref{eq:P-constraint}), to define the widths $h_i$ for $Nd>1$. Again, the recurrence relation \eqref{eq:bandwidths} can be employed to quickly obtain an
approximate solution. 

\begin{figure*}
    \begin{center}
        \includegraphics{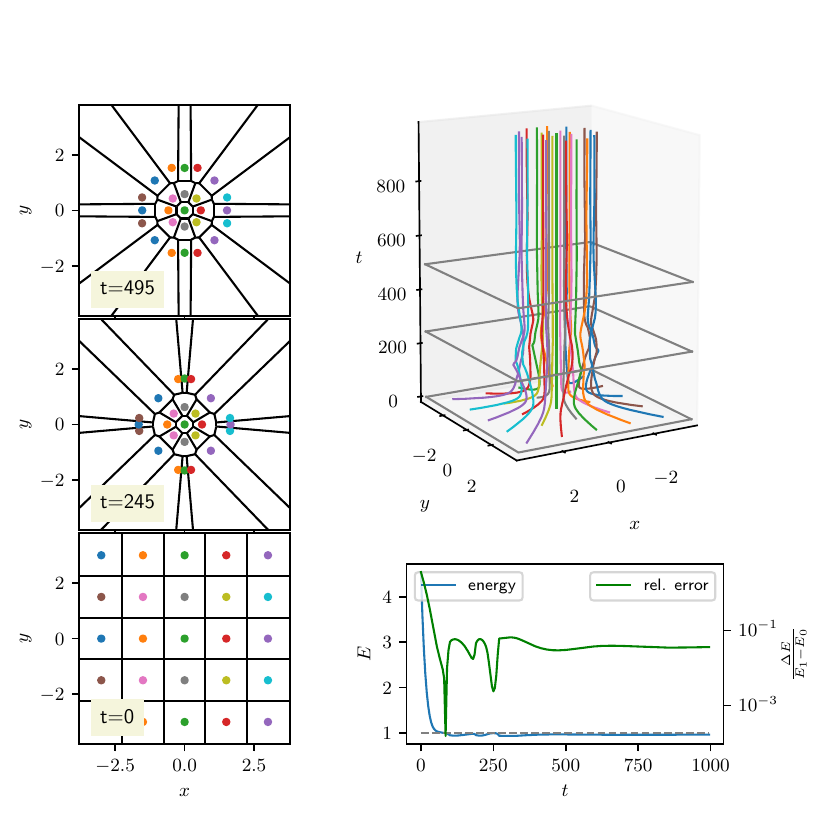}
        \caption{
            \label{fig:harmonic}
            Convergence of the Gaussian kernel algorithm for $M=25$ worlds,
            $N=1$ particle, and $d=2$ spatial dimensions, in a harmonic
            potential $V(X)=\half\hbar^2\omega^2 X\tp X$. \blk The plot on the upper
            right shows the evolution of the configuration of worlds ${\cal
            Q}$ during the iteration of the algorithm. The integration time
            step was chosen to be $\Delta t= 5\cdot 10^{-2}$ for $2\cdot 10^4$
            integration steps. The three plots on the left are 
            configuration space snapshots for the respective times $t$ shown in
            the lower left corner, respectively. The circles denote the worlds
            and the black lines illustrate their respective Voronoi cells. The
            plot in the lower right displays the convergence of the energy,
            again in units of $\hbar\omega$, as well as the relative error. The
            latter is computed as ratio of the difference of the difference
            w.r.t.\ the exact ground state energy (dashed line) and the difference of the
            exact energies of the first excited and the ground state. Note that
            in 2d the excited and ground state energies are given by
            $E_1=2\hbar\omega$, and $E_0=\hbar\omega$.
        }
    \end{center}
\end{figure*}

\begin{figure*}
    \begin{center}
        \includegraphics{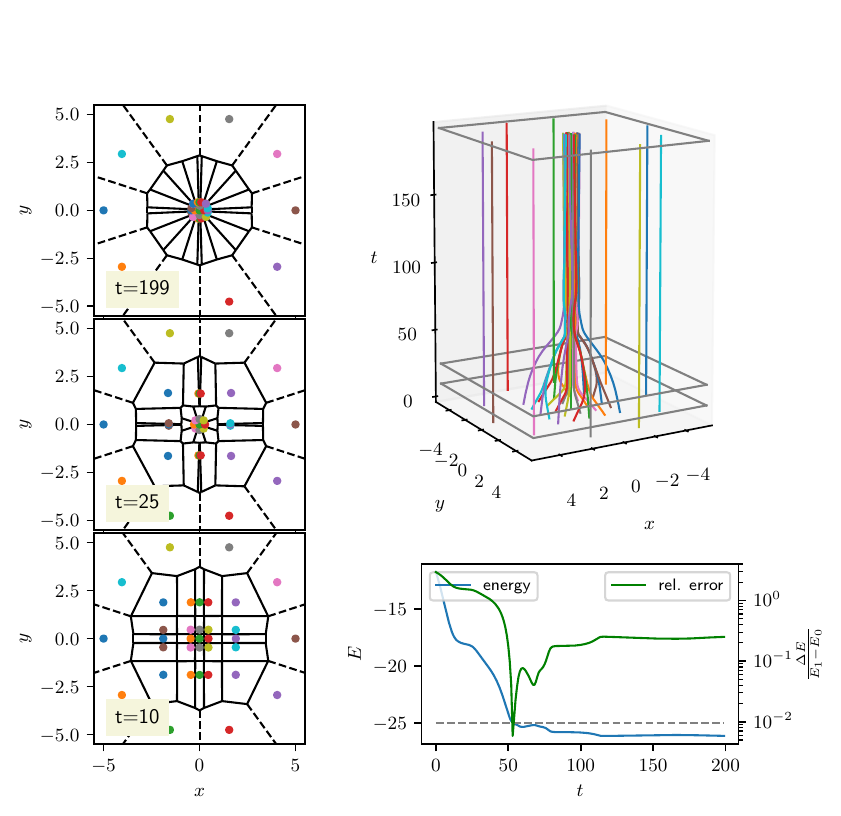}
        \caption{ \label{fig:poeschl} 
            Convergence of the Gaussian kernel algorithm for $M=25$ worlds, $N=1$ particle, and $d=2$ spatial
            dimensions, in a Pöschl-Teller type potential defined as $V(X)=V_1(x_1)+V_1(x_2)$ for $X=(x_1,x_2)$ and
            $V_1(x)=\frac{\alpha^2}{2}\frac{\lambda(\lambda+1)}{\rm cosh^2(\alpha x)}$, where we chose $\lambda=5$. The
            plot on the upper right shows the evolution of the configuration of worlds ${\cal Q}$ during the iteration
            of the algorithm. The integration time step was chosen to be $\Delta t= 2\cdot 10^{-2}$ for $10^4$
            integration steps. The three plots on the left are configuration space snapshots for the respective times
            $t$ shown in the lower left corner, respectively. The circles denote the worlds and the black lines
            illustrate their respective Voronoi cells. Note the boundary worlds which, as discussed in the text, have
            been fixed to lie on a circle of sufficiently large radius to stabilize the iteration. The plot in the lower
            right displays the convergence of the energy, again in units of $\alpha^2\hbar^2/m$, as well as the relative
            error. The latter is computed as ratio of the difference of the difference w.r.t.\ the exact ground state
            energy (dashed line) and the difference of the exact energies of the first excited and the ground state.
            Note that in 2d the excited and ground state energies are given by $E_1=-20.5\cdot\alpha^2\hbar^2/m$, and
            $E_0=-25\cdot\alpha^2\hbar^2/m$. The computed energy lies systematically below the exact ground state
            energy. As discussed, this systematic error is caused by the artificially fixed boundary worlds.
        }
    \end{center}
\end{figure*}

We have tested this generalized Gaussian kernel algorithm for $N=1$ particles
and $d=2$ spatial dimensions, in two cases: 1) A harmonic potential
$V(X)=\half\hbar^2\omega^2 X\tp X$ and 2) a Pöschl-Teller type potential. Note
that, unlike the harmonic potential, the Pöschl-Teller potential has many more
or less natural generalizations in more than one spatial dimension. For our
proof of concept study we took the simple choice: $V(X)=V_1(x_1)+V_1(x_2)$ for
$X=(x_1,x_2)$, $V_1(x)=\frac{\alpha^2}{2}\frac{\lambda(\lambda+1)}{\rm
cosh^2(\alpha x)}$ and $\lambda=5$.
 The results of the corresponding numerical simulations for
$M=25$ worlds are shown in figures~\ref{fig:harmonic} and \ref{fig:poeschl},
respectively, showing convergence to corresponding ground state configurations.

As discussed above, the non-crossing property of worlds is not an issue for
$Nd>1$. However, one has to consider a potentially more serious problem
concerning the boundary worlds in the outer cells, for which the {{\it a
priori}} distribution in Eq.~(\ref{eq:genP-constraint}) reduces to an
uninformative value of $p_i=0$, independently of the actual positions of the
boundary worlds. In the case $Nd=1$ there are only two boundary worlds
$Q^{({1})}$ and $Q^{(M)}$ whereas, e.g., in our setup for the harmonic
potential, with $Nd=2$ and $M=25$, we have 16 boundary worlds as depicted in
figure~\ref{fig:harmonic}. 


In the case of the harmonic potential the boundary worlds were not found to be
problematic, essentially because while their motion is not moderated by other
worlds, the strongly-confining nature of the potential does not allow any of
the worlds to escape to spatial infinity. In contrast, the Pöschl-Teller
potential is asymptotically constant, and hence does not confine the boundary
worlds. Due to that fact the corresponding numerical simulation easily becomes
unstable. However, a straight-forward solution to circumvent this problem is to
introduce additional artificial worlds at fixed positions surrounding the
actual worlds ${\cal  Q}$; see the straight  trajectories plotted \blk in 
figure~\ref{fig:poeschl}.  These artificial boundary worlds act to 
damp any unwanted oscillations of the outer worlds of ${\cal  Q}$ but on the
other encode a kind of boundary condition on the Hamiltonian at hand. Hence,
these boundary worlds must be placed 
at sensible locations, having a sufficient distance
to the actual worlds ${\cal  Q}$, so that the accuracy of the world distribution is
only changed in regions of configuration space where the density should in any
case be very low. One may therefore expect that the accuracy of the numerically
inferred moments of observables are not significantly affected. We have not
tried to optimize the location of the boundary worlds in our first trial in
figure~\ref{fig:poeschl}, which is why the numerically determined value of the
ground state is systematically smaller that the exact one.

All these troubles seem to be connected to the discrepancy between the two required approximation modes, i.e., in the
$L^2$-norm sense, required for the statistics, and the one in the point-wise sense, required to obtain the world
trajectories. Our choices made above in terms of subdivisioning methods, approximation kernels and their corresponding
parameters, can be seen as forms of relieving this discrepancy through specification of {\it{a priori}} knowledge about
the smoothness. These phenomena would of course have to be studied in more detail, however, our analysis already
indicates that also in more than one spatial dimension one may expect our proposed approach to be applicable to ground
and excited states for various potentials.
 
\section{Conclusions}
\label{sec:conclusion}

Although there are still several issues to address in order to arrive at a robust integrator for very general settings,
we were able to provide a generalization of the MIW algorithm~\cite{prx}, which previously \blk could only treat the
case of one particle in one spatial dimension.  This generalization was shown to perform well for calculating ground
state energies and configurations for the harmonic and Pöschl-Teller potentials, showing good quantitative agreement
with the exact solutions in one and two spatial dimensions. We furthermore, demonstrated in one spatial dimension that
the proposed algorithm can be adapted to find excited states provided the position of the nodes is known \emph{a
priori}. It is very likely that this algorithm can also be generalized in a way that does not assumed the given
positions of the nodes but only a given number of them. The corresponding positions of the nodes may then be found by
minimizing the energy functional using a gradient descent. It should, however, be emphasized that in more than one
spatial dimension the point-like nodes become nodal surfaces, which render the problem of finding excited states more
complicated.  \blk

Very recently, kernel density estimators have also been applied by Sturniolo, within the framework of the MIW approach,
to similarly smooth the empirical density, with numerical calculations made for ground states of harmonic and
Lennard-Jones potentials in two and three dimensions~\cite{sturn}, using a different method for constructing the
estimator,  again with promising results. Sturniolo further suggests that exponential kernels may perform better than
Gaussian kernels  for finding ground state energies, but worse for finding ground state configurations, and that it may
be possible to simulate temperature-dependent tunneling effects in the MIW approach~\cite{sturn}.

The original motivation in using the MIW approach in numerical computations was the hope of a generally-applicable
method that reduced computational resources as compared to fixed-grid methods. At first sight, the provided numerical
implementation still seems to be computationally expensive. Even when neglecting the iteration that determines the
bandwidths $h_i$ in each integration step, see~\eqref{eq:bandwidths}, the computational effort scales at least as $M^2$,
as $M$ contributions to the potential must be calculated for each world via Eq.~(\ref{eq:estimator}). There are,
however, many tricks to reduce this scaling.  Foremost, it has been claimed~\cite{elgmmal}, with respect to kernel
density estimation approaches, that in general situation this scaling can be reduced to a linear one in $M$.  The
leading idea behind such an improvement is based on the fact that due to the choice of bandwiths $h_i$ the Gaussian
kernel functions are usually highly peaked in regions where many worlds cluster. Hence, the corresponding kernel
functions fall of rapidly and the sum in the density estimator \eqref{eq:estimator} can be truncated. The computational
effort to compute values of the exponential function can be reduced further by replacing it with a fixed lookup table
that is interpolated according to the scaling introduced by $h_i$. Finally, it also has to be emphasized that the
iteration to determine the $h_i$ in \eqref{eq:bandwidths} usually converges sufficiently after very few iterations,
since the configurations of the worlds ${\cal  Q}$ change only slightly between the integration steps. However, it is
also conceivable that the $h_i$ can be determined dynamically from the \blk world configurations at each time step.
Recently ideas were explored in Ref.~\cite{dispersion} to determine the bandwiths dynamically, by comparison with
dispersion ruled by the heat equation.  It seems feasible to obtain similar dynamical laws for the $h_i$ when the
dispersion is ruled by the Schrödinger equation.

The crucial next step should be a systematic study of the $M$-dependent scaling of the proposed algorithm given a fixed
numerical accuracy that has to be met. Such a study should decide whether the algorithm lives up to the expectation that
the exponential scaling of fixed-grid methods can be avoided while maintaining the same numerical accuracy. Thus, we
have reasonable confidence that development of our approach in this paper will lead to a general and efficient numerical
tool for ground state and other calculations. \blk

\paragraph{Acknowledgments} We would like to thank  M. Ghadimi  and T. Gould \blk for valuable discussions. Furthermore,
D.-A.D.\ and H.H.\ would like to thank Griffith University for its hospitality,  while M.J.W.H.\ and H.M.W.\  likewise
thank the Ludwig Maximilian University. \blk This work was partially funded by the Elite Network of Bavaria through the
Junior Research Group “Interaction between Light and Matter” and by FQXi Grant FQXi-RFP-1519. \blk



\begin{thebibliography}{99}

\bibitem{lieb} E.H. Lieb, R. Seiringer, J.P. Solovej, and J. Yngvason,
{\it The Mathematics of the Bose Gas and its Condensation.}
Oberwolfach Seminars, Vol. 34, Birkh\"auser (2005)
\bibitem{wyatt}  R. E. Wyatt, {\it Quantum Dynamics with Trajectories} (Springer, New York, 2005).

\bibitem{chatteraj} P. K. Chatteraj, {\it Quantum Trajectories} (CRC Press, 2010).

\bibitem{deckert} D.-A. Deckert, D. D\"urr, and P. Pickl. {\it Quantum Dynamics
    with Bohmian Trajectories}, J. Phys. Chem. A {\bf 111}, 10325 (2007).

\bibitem{detlef} D. Duerr, S. Teufel, {\it Bohmian Mechanics. The Physics and
Mathematics of Quantum Theory.}, Heidelberg. Springer (2009).

\bibitem{artes} Sanz, Ángel S., and Salvador Miret-Artés. {\it Quantum
    Mechanics with Trajectories.} in {A Trajectory Description of Quantum
Processes. I. Fundamentals}, 187–230. Lecture Notes in Physics.
Springer, Berlin, Heidelberg (2012).

\bibitem{sanz} Benseny, Albert, Guillermo Albareda, Ángel S. Sanz, Jordi
Mompart, and Xavier Oriols. {\it Applied Bohmian Mechanics.} The European
Physical Journal D 68 (10):286 (2014). 

\bibitem{prx} M. J. W. Hall, D.-A. Deckert, and H. M. Wiseman, {\it Quantum
Phenomena Modeled by Interactions between Many Classical Worlds}, Phys. Rev. X
4, 041013 (2014).
\bibitem{GhaHalWis18}
M. Ghadimi, M. J.W. Hall, H. M. Wiseman,
{\it Nonlocality in Bell's Theorem, in Bohm's Theory, and in Many Interacting Worlds Theorising},
Entropy 20,
567 (2018).

\bibitem{holland} P. Holland, {\it Computing the Wavefunction from
Trajectories: Particle and Wave Pictures
in Quantum Mechanics and Their Relation}, Ann. Phys. {\bf 315}, 505 (2005).
\bibitem{poirier} B. Poirier, {\it Bohmian Mechanics without Pilot Waves},  Chem. Phys. {\bf 370}, 4 (2010).
\bibitem{parlant} G. Parlant, Y.C. Ou, K. Park and B. Poirier, {\it Classical-Like Trajectory Simulations for Accurate Computation of Quantum Reactive Scattering Probabilities},
  Computat. Theoret. Chem. {\bf 990}, 3 (2012).
\bibitem{schiff} J. Schiff and B. Poirier, {\it Quantum Mechanics without
Wavefunctions}, J. Chem. Phys. {\bf 136}, 031102 (2012).

\bibitem{sebens} C. Sebens, {\it Quantum mechanics as classical physics},
Philosophy of Science Vol. 82, No. 2, pp. 266-291 (2015)
\bibitem{bostrom} K. J. Bostr\"om, {\it Quantum mechanics as a deterministic theory of a continuum of worlds}, Quantum Stud.: Math. Found. 2, 315 (2015).
\bibitem{smolin} L. Smolin, {\it Quantum mechanics and the principle of maximal variety}, Found. Phys. 46, 736 (2016).
\bibitem{roser} P. Roser, M. T. Scoggins,
{\it Non-Quantum Behaviors of Configuration-Space Density Formulations of quantum mechanics}, Eprint, arXiv:2303.04959

\bibitem{BerZan05} K. Berndl, M. Daumer, D. D\"urr, S. Goldstein and N. Zangh\`i,  {\it A Survey of Bohmian Mechanics}, Il Nuovo Cimento B
(1971-1996) {\bf 110}, 737 (1995). 

\bibitem{mckeague} I. W.  McKeague and B. Levin, {\it Convergence of empirical distributions in an interpretation of quantum mechanics}, Ann. Appl. Probab. {\bf 26}, 2540-2555 (2016).
 
\bibitem{hannes} H. Herrmann, {\it Finding stationary states by interacting many worlds}, Master Thesis, Mathematical
    Institute of the LMU Munich (2016).

\bibitem{git} H. Herrmann and D.-A. Deckert, {\it Eigenstates in the Many Interacting Worlds approach}, GitLab repository, \url{https://gitlab.com/dirk-deckert-lmu/eigenstates-in-the-Many-Interacting-Worlds-approach} (2023).

\bibitem{sturn} S. Sturniolo, {\it Computational applications of the many-interacting-worlds interpretation of quantum mechanics}, Phys. Rev. E 97, 053311 (2018). \blk

\bibitem{izenman} A. J. Izenman, {\it Review Papers: Recent Developments in
Nonparametric Density Estimation}, Journal of the American Statistical
Association, 86, 205–224 (1991).
\bibitem{scott} W. Scott, {\it Multivariate density estimation: theory,
practice, and visualization}, Hoboken, New Jersey: John Wiley \& Sons, Inc,
second edition ed. (2015).
\bibitem{abramson} I. S. Abramson, {\it On Bandwidth Variation in Kernel
Estimates-A Square Root Law},  The Annals of Statistics, 10, 1217–1223 (1982).
\bibitem{elgmmal} A. Elgammal, R. Duraiswami, and L. S. Davis, {\it Efficient Kernel Density
Estimation using the Fast Gauss Transform with Applications to Color Modeling
and Tracking}, IEEE Transactions on Pattern Analysis and Machine Intelligence,
25, 1499–1504 (2003).

\bibitem{delaunay} J. A. De Loera, J. Rambaau and F. Santos, {\it Triangulations}, vol. 25 of {\it Algorithms and Computation in Mathematics} (Springer, Berlin, 2010).
\bibitem{dispersion} Z. I. Botev, J. F. Grotowski and D. P. Kroese, {\it Kernel density estimation via diffusion}, The Annals
of Statistics,
Vol. 38, No. 5, 2916–2957, 2010.
\bibitem{courant} R. Courant and D. Hilbert, {\it Methoden der mathematischen
Physik.} Springer-Verlag, 1924.
\bibitem{aconda} A. Ancona, B. Helffer, and T. Hoffmann-Ostenhof, {\it Nodal
domain theorems a la Courant}, Documenta Mathematica, vol. 9, pp. 283–299,
2004.


\end{thebibliography}
\end{document}